\documentclass[aps,pre,twocolumn,amsmath,amssymb,superscriptaddress,preprintnumbers,showpacs,floatfix,pdftex]{revtex4-1}

\usepackage[pdftex]{graphicx}
\usepackage{dcolumn}
\usepackage{multirow}

\newcommand{\ie}{\emph{i.e., }}
\newcommand{\cf}{\emph{cf. }}
\newcommand{\eq}[1]{(\ref{#1})}
\newcommand{\X}{\ensuremath{\bullet}}
\renewcommand{\O}{\ensuremath{\circ}}
\newcommand{\rmd}{\ensuremath{\textrm{d}}}

\begin{document}
%IOP
\title[Network representations of NESS: Cycles, symmetries and dominant paths]
{Network representations of non-equilibrium steady states:\\ 
  Cycle decompositions, symmetries and dominant paths}

\author{B Altaner}
\affiliation{Max-Planck-Institute for Dynamics and Self-Organization, G\"ottingen, Germany}
\affiliation{Faculty of Physics, Georg-August Univ.~G\"ottingen, G\"ottingen, Germany} 

\author{S Grosskinsky} 
\affiliation{Mathematics Institute, University of Warwick, Coventry, UK}

\author{S Herminghaus}
\affiliation{Max-Planck-Institute for Dynamics and
  Self-Organization, G\"ottingen, Germany}
\affiliation{Faculty of Physics, Georg-August Univ.~G\"ottingen, G\"ottingen, Germany} 

\author{L Katth\"an}
\affiliation{FB Mathematik und Informatik, Philipps Univ.~Marburg, Marburg, Germany}

\author{M Timme}
\affiliation{Max-Planck-Institute for Dynamics and
  Self-Organization, G\"ottingen, Germany}
\affiliation{Faculty of Physics, Georg-August Univ.~G\"ottingen, G\"ottingen, Germany} 

\author{J Vollmer}
\affiliation{Max-Planck-Institute for Dynamics and
  Self-Organization, G\"ottingen, Germany}
\affiliation{Faculty of Physics, Georg-August Univ.~G\"ottingen, G\"ottingen, Germany}

\begin{abstract}
  Non-equilibrium steady states (NESS) of Markov processes give rise
  to non-trivial cyclic probability fluxes.
  % MT;out: as a result of probability conservation.
  Cycle decompositions of the steady state offer an effective
  description of such fluxes.  Here, we present an iterative cycle
  decomposition exhibiting a natural dynamics on the space of cycles
  that satisfies {\it detailed balance}.  Expectation values of
  observables can be expressed as cycle ``averages'', resembling the
  cycle representation of expectation values in dynamical systems.  We
  illustrate our approach in terms of an analogy to a simple model of
  mass transit dynamics. Symmetries are reflected in our approach by a
  reduction of the minimal number of cycles needed in the
  decomposition.  These features are demonstrated by discussing a
  variant of an asymmetric exclusion process (TASEP).  Intriguingly, a
  continuous change of dominant flow paths in the network results in a
  change of the structure of cycles as well as in discontinuous jumps
  in cycle weights.
\end{abstract}

\pacs{05.70.Ln, 05.10.Gg, 89.75.Fb}

\maketitle

%% -------------------------------------- %%
\section{Introduction}

A major challenge of statistical physics is to identify principles
organizing the structure of steady states \cite{Haken1983}.
Equilibrium systems are singled out by detailed balance, a symmetry in
the transition rates between different states that explicitly yields
the systems' free energies \cite{Schnakenberg1976,Zia+Schmittmann2007}
and thereby all its linear thermodynamic properties. In
non-equilibrium steady states (NESS), detailed balance is broken and
non-trivial currents can be identified.

Following Penrose \cite{Penrose1970}, we idealize observable processes as irreducible Markov processes on a finite state space.
Here, irreducible means that the system can reach any state \(i\) from any other state \(j\) with a finite number of transitions.
%Recurrence means that in an infinite time span the system visits every state infinitely often.
On a finite state space this implies ergodicity and hence ensures the existence of a steady state \cite{Feller1968}.

Conservation of probability in the form of Kirchhoff's law induces probability flux cycles \cite{Schnakenberg1976,Zia+Schmittmann2007,Kalpazidou1993, Kalpazidou2006, Andrieux+Gaspard2007, Faggionato+dPietro2011, Jiang_etal2004,Hill1977}, and there are a number of distinct ways to decompose the stationary dynamics as cycles:
%Here, we present a method to map NESS fluxes onto a Markov process on a dual space of flux cycles.
%Detailed balance is restored in that description, which allows to define a potential function as in the case of equilibrium systems.
%Steady-state averages take the form of equilibrium averages on that dual space.
%THIS WAS MOVED A BIT DOWNWARDS, AFTER "... Hill's cycles"
%The idea of representing NESS in terms of cycles is not new. 
Schnakenberg network theory (SNT, \cite{Schnakenberg1976}) and subsequent work (see eg.~\cite{Zia+Schmittmann2007,Andrieux+Gaspard2007}) is based on identifying a fundamental set of cycles after identifying a spanning tree.
A recent approach further generalizes those results to a different basis of oriented cycles \cite{Faggionato+dPietro2011}.
From a more mathematical point of view, Kalpazidou \cite{Kalpazidou2006} and the Beijing school of Quians \cite{Jiang_etal2004} independently developed a rigorous formalism to describe Markov processes on finite (and countably infinite) state spaces using cycles.
They distinguish between {\it stochastic} and {\it deterministic} decomposition algorithms.
The former leads to the cycle decompositions used by Hill \cite{Hill1977} and has an effective dual description as a Markov process on the set of all possible cycles.
It also has the benefit that a cycle can be interpreted as the so-called {\it completion rate} of this cycle within the stochastic dynamics \cite{Kalpazidou2006,Jiang_etal2004}.
The latter, deterministic approach is closely related to the algorithm used in the present work.
It is complementary to both SNT and Hill's cycles.

Here, we present a method to map NESS fluxes onto a Markov process on a dual space of flux cycles.
Detailed balance is restored in that description, which allows to define a potential function as in the case of equilibrium systems.
Steady-state averages take the form of equilibrium averages on the dual space.

\begin{figure*}
\begin{center}
 \includegraphics[width=0.85\textwidth]{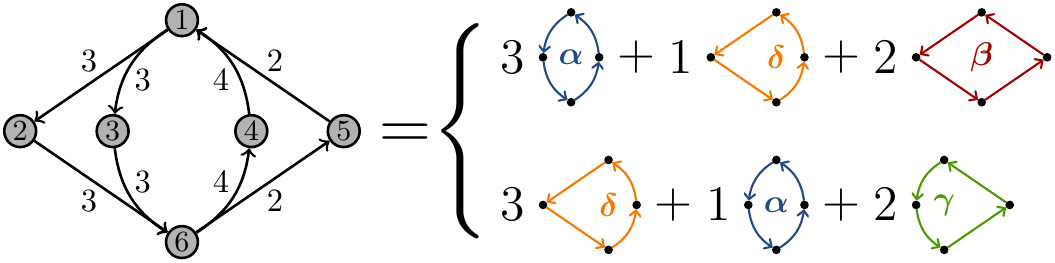} \vspace{-.015\textwidth} 
\end{center}
 \caption{Representation of a NESS in terms of linear superpositions of cycle fluxes.
 The numbers on the arrows (representing the directed transitions) are the values of the fluxes.
 The steady-state fluxes between the states \textcircled{\tiny 1} - \textcircled{\tiny 6} can be decomposed into cycle fluxes (labeled by Greek letters) with positive weights.
  Two different decompositions are possible.
  \label{fig:markov-graph}}
\end{figure*}

The essence of our approach is best viewed in the ensemble picture.
Consider a large number of identical physical systems with a finite number of states.
Each system entering a certain state \(i\) stays there for an average time
 \(\langle \tau_i\rangle\),
and then proceeds to another state \(j\) according to a fixed transition rate.
Up to normalization the flux may be seen as the number of systems proceeding from one state to another per unit time.
In figure~\ref{fig:markov-graph} we present an elementary six-state example motivated by the TASEP example discussed below (\cf figure~\ref{fig:TASEP-example} and tables \ref{tab:TASEP-states}-\ref{tab:TASEP-decompositions}).
The cycle representation of the fluxes means to write them as a linear superposition of cycle fluxes with a non-negative weight assigned to each cycle.
Such representations exist for any NESS \cite{Kalpazidou1993,Kalpazidou2006}.

%To better follow the line of arguments, it is helpful to consider a socio-physical analogy:
%the cycles may be interpreted as the lines of a mass transit system with the peculiarity that the lines are running one-way on closed loops.
%The fluxes are proportional to the total amount of passengers travelling from one station to another; \ie from a state \(i\) to a state \(j\) of the Markov process.
%The lines are represented in different colors in figure~\ref{fig:markov-graph}.
%We imagine each passenger to carry a (correspondingly colored) ticket indicating the line he is currently using.
%Passengers can change lines in the stations.
%To remain in a steady state this involves a random exchange of tickets between passengers at stations.
%THIS WAS MOVED TO THE BEGINNING OF SECTION II

The aim of this work is to explore consequences of this point of view on NESS, with emphasis on 
%
%Mathematical questions concern the existence and uniqueness of such a representation.
%Figure~\ref{fig:markov-graph} shows by example that a representation in terms of a linear superposition of cycles is not unique.
%The question concerning existence will be answered positively in the following.
%Physically more interesting is the question for 
the relation of cycles and non-equilibrium phase-transitions.

The paper is organized as follows.
In sections \ref{sec:mp-revisited}--\ref{sec:algo-proofs} we introduce cycles as a topological backbone of NESS.
%To that end, in parts we revisit mathematical literature (\cf\cite{Kalpazidou2006,Jiang_etal2004}) on that issue.
In section \ref{sec:numbers} we compare the present approach to other physical cycle theories, \cite{Schnakenberg1976,Hill1977}.
We see that a natural stochastic dynamics leads to detailed balance on the dual space of cycles in section \ref{sec:detbal}.
Section \ref{sec:averages} defines Boltzmann-like averages on cycle space.
They are related to physical current variables in section~\ref{sec:physics}.
Finally, in section \ref{sec:example} we use a TASEP as an example to investigate how phase-transitions and symmetries influence the cycle structure.
Appendix \ref{sec:analogy} describes useful thermodynamic and electric analogies which also remain valid in the discrete case, as explained in appendix \ref{sec:discrete}.

%% -------------------------------------- %%
%% -------------------------------------- %%

\section{Markov processes revisited}
\label{sec:mp-revisited}

We start by briefly reviewing Markov processes on a finite state space \cite{Feller1968,Schnakenberg1976,Zia+Schmittmann2007}. 
To better follow the line of arguments, it is helpful to consider a socio-physical analogy:
the cycles may be interpreted as the lines of a mass transit system with the peculiarity that the lines are running one-way on closed loops.
The fluxes are proportional to the total amount of passengers traveling from one station to another; \ie from a state \(i\) to a state \(j\) of the Markov process.
The lines are represented in different colors in figure~\ref{fig:markov-graph}.
We imagine each passenger to carry a (correspondingly colored) ticket indicating the line he is currently using.
Passengers can change lines in the stations.
To remain in a steady state this involves a random exchange of tickets between passengers at stations.

We represent the process as a random walk on a graph \(G=(V,E)\) with \(N=\vert V \vert\) vertices \(v_i\), \(i \in \left\{ 1,\ldots, N\right\} \) and directed edges \((i,j) \in E\).
The vertices represent the states of the system, and are shown as gray circles displaying the vertex indices \(1,\ldots,6\) in figure ~\ref{fig:markov-graph}.
A system entering vertex \(v_i\) will jump to another vertex \(v_j\) with probability \(a^i_j\) after having stayed in state \(i\) for an exponentially distributed waiting time \(\tau_i\).
Consequently, the (time-independent) transition rates per unit time are
 \(w^i_j := a^i_j/\langle \tau_i \rangle\).
A system trajectory is the realization of a random walk of one of the passengers through the transit system.
In terms of the transition matrix
\begin{equation}
 W^i_j :=\begin{cases}
    w^i_j& \textrm{for } i\neq j\\ 
   -\langle \tau_i \rangle^{-1} \equiv -\sum_{k\neq i} w^i_k  & \textrm{for } i = j\\
 \end{cases}
 \label{eq:transition-matrix}
\end{equation}
or for the fluxes from \(i\) to \(j\neq i\)
\begin{equation}
 \phi^i_j(t) = p_i(t) w^i_j
 \label{eq:flux-defn}
\end{equation}
the equation for the evolution of the probability \(p_i(t)\) to find the system in a state \(i\) at time \(t\) takes the compact form
\begin{equation}
 \frac {\rmd p_i}{\rmd t} = \sum_j W^j_i p_j = \sum_{j\neq i} \left( \phi^j_i - \phi^i_j\right).
 \label{eq:master-equation}
\end{equation}
Here and in the following we suppress the explicit time-dependence and write, e.g., \(p_i\) instead of \(p_i(t)\).
The first equality in (\ref{eq:master-equation}) stresses the linearity of the problem and is useful for algebraic considerations.
The second emphasizes the physical concept of a master or continuity equation: in a steady state the net influx must equal the net outflux,
\(\sum_{j\neq i}{\phi^i_j}^* = \sum_{j\neq i}{\phi^j_i}^*\).
In terms of the currents,
 \(I^i_j := \phi^i_j - \phi^j_i\),
this node condition, 
\begin{equation}
  0 \stackrel{!}{=} \sum_{j\neq i}\left( {\phi^i_j}^* - {\phi^j_i}^*\right) = \sum_{j\neq i} {I^i_j}^*,
 \label{eq:node-condition}
\end{equation}
amounts to Kirchhoff's current law which expresses particle (or probability) conservation at each vertex \cite{Kirchhoff1847}.
Here and in the following the~\(^{\ast}\) marks steady-state quantities.

Due to the continuity equation (\ref{eq:master-equation}) every normalized initial distribution remains normalized at all times, and it relaxes to a steady state \(p^*_i\) \cite{Schnakenberg1976}.

Algebraically the steady-state probability distribution \(p_i^{\ast}\) is a left eigenvector of \(W\) with eigenvalue zero.
Ergodicity ascertains the existence of a path \(i_0 \ldots i_n\) with a positive
 \(\omega_{i_0,\ldots,i_n} := \prod_{j=1}^{n} w^{i_{j-1}}_{i_{j}}\)
for every pair of vertices \(i_0\) and \(i_n\).
This ensures existence and uniqueness of the normalized distribution obeying
\begin{equation}
 \sum_i p^*_i = 1.
 \label{eq:normalize-p}
\end{equation}

In the physics literature a steady state is called an equilibrium if it obeys {\it detailed balance}, \ie if the individual fluxes between any two vertices \(i\) and \(j\) cancel, \ie
\begin{equation}
 {I^i_j}^*={\phi^i_j}^* - {\phi^j_i}^* =0 .%, ~\forall i,j .
 \label{eq:det-balance}
\end{equation}

Detailed balance further implies a weaker symmetry that is sometimes called {\it dynamical reversibility} \cite{Maes2004}.
It means that if a transition is allowed, so is its reverse, \ie
\begin{equation}
  w^i_j > 0 \Leftrightarrow w^j_i > 0.
  \label{eq:dynamical_reversibility}
\end{equation}
The cycle decomposition does not need this symmetry in the transition rates, \ie we allow for unidirectional transitions.
Consequences of dynamically reversible systems are discussed below.

For an equilibrium system the ratio of \(\omega_{i_0,\ldots,i_n}\)
and the one for the reverse path \(\omega_{i_n,\ldots,i_0}\)
only depends on the initial and final point irrespective of the chosen path \cite{Zia+Schmittmann2007,Schnakenberg1976}.
Examining the above relation for paths starting from a fixed reference vertex \(j\) one obtains an explicit representation of the steady-state probability density
\begin{equation}
 p_i^{\ast} = p_j^{\ast} \;\; \frac{\omega_{j,\ldots,i}}{\omega_{i,\ldots,j}}
 =: p_j^{\ast} \; \exp\left( - U^{(j)}_i \right) .
 \label{eq:Potential-DB}
\end{equation}
Then one can always write \(U^{(j)}_i = U_i+ c_j\),
where \(U_i\) is a universal function and \(c_j\) depends on the chosen reference site.
Consequently, 
\begin{subequations}
 \begin{equation}
  p_i^{\ast} = Z^{-1} \exp( - U_i )
 \end{equation}
 where the partition function 
 \begin{equation}
  Z = \sum_k \exp( - U_k )
 \end{equation}%
\end{subequations}
secures normalization, (\ref{eq:normalize-p}).

%% -------------------------------------- %%
\section{Cycle representation and transform}
\label{sec:transform}
The cycle transform is based on the idea that fluxes in a steady state may be represented as superpositions of cycle fluxes (cf.\ figure \ref{fig:markov-graph}).
A cycle \(\alpha\) of length \(s_{\alpha}\) is an equivalence class of ordered sets of \(s_{\alpha}\) vertices which form a self-avoiding closed path, where paths differing only by a cyclic permutation of vertices are identified.
We quantify the number of systems traversing each edge of \(\alpha\) by the weight \(m_\alpha^{\ast}\).
There can be several cycles along an edge \((i,j)\) and the flux \(\phi^i_j\) quantifies the {\it total} number of systems traversing that edge per unit time.
In the remainder of this section we work out how the steady-state fluxes can be represented by different cycles \(\alpha\) with positive weights \(m^*_{\alpha}\) assigned to each of them.

To express the geometrical structure of the cycles we define the indicator (or passage) functions 
\(\chi^i_{j,\alpha}\) and \(\chi_{i,\alpha} \) as
\begin{subequations}
 \begin{eqnarray}
   \chi^i_{j,\alpha}  &= &
   \begin{cases}
     1  \textrm{ if } \alpha \textrm{ passes through the directed edge } (i,j)\\ 
     0  \textrm{ otherwise}
   \end{cases}\\
   \chi_{i,\alpha}  &= &
   \begin{cases} 
     1  \textrm{ if } \alpha \textrm{ passes through vertex } i\\ 
     0  \textrm{ otherwise}
   \end{cases}
 \end{eqnarray}
   \label{eq:pi-defn}
\end{subequations}
In the language of graph theory \(\chi^i_{j,\alpha}\) is the adjacency matrix of a cycle.
The following identities hold:
\begin{subequations}
  \begin{eqnarray}
   \sum_j \chi^i_{j,\alpha} 
   &=&
   \sum_j \chi^j_{i,\alpha} = \chi_{i,\alpha}, \label{eq:geometry-identity}
   \\
   \sum_i \chi_{i,\alpha} 
   &=&
   s_\alpha,
  \end{eqnarray}
\end{subequations}
  where \(s_{\alpha}\) is the length of the cycle \(\alpha\).
  With their help we formulate the ideas of the previous paragraph mathematically.
  As we show below, there is a set of cycles $\{ \alpha_k \}$ with non-negative flux densities \(m_\alpha^{\ast}\geq 0\) such that
  \begin{equation}
   {\phi^i_j}^{\ast}  =  \sum_{\alpha} m_\alpha^{\ast}\chi^i_{j,\alpha} \,  \label{eq:weight-defn}
  \end{equation}
  for all pairs of vertices \((i,j)\).

  To obtain a decomposition we choose an arbitrary enumeration of all \(M\) possible cycles \(\alpha_1\), \(\alpha_2\), \ldots , \(\alpha_M\) on $G$.
  The ambiguity in choosing the order of this enumeration leads to different decompositions constructed by the following algorithm: \\
  Start the iteration for cycle \(\alpha_1\) with a flux field
   \({\phi^i_j}^{(1)} = {\phi^i_j}^\ast\)
  that contains the steady-state fluxes of the original system:
  \begin{itemize}
  \item
   Initialization:
   \begin{equation}
     {\phi^i_j}^{(1)} := {\phi^i_j}^*, ~\mathrm{for~all}~ i,j.
    \\
    \label{eq:algo-initialize}
   \end{equation}
  \end{itemize}
  Successively subtract the fluxes along different cycles.
  In the \(k\)th step set \(m^*_{\alpha_k}\) to be the minimum of the values of the flux \( {\phi^i_j}^{(k)}\) along the edges contained in \(\alpha_k\).
  The new flux field in iteration \(k+1\) is the current one with \(m^*_{\alpha_k}\) subtracted at the edges traversed by cycle \(\alpha_k\):
  \begin{itemize}
  \item
   Iteration:
\begin{subequations}
    \begin{eqnarray}
     m_{\alpha_{k}}^\ast &%
     :=&
     \min_{i,j}\{{\phi^i_j}^{(k)} : {\chi^i_{j,\alpha_{k}}} > 0 \}, 
     \label{eq:assigned-m}\\
     {\phi^i_j}^{(k+1)} &%
     :=&
     {\phi^i_j}^{(k)} - m_{\alpha_{k}}^\ast {\chi^i_{j\alpha_{k}}}.
     \label{eq:assigned-phi}
    \end{eqnarray}
    \label{eq:algo-iterate}
\end{subequations}
  \end{itemize}
  The algorithm terminates after all possible cycles have been considered:
  \begin{itemize}
  \item
   Termination condition:\\
   \begin{equation}
    k = M
    \label{eq:algo-terminate-cond}
   \end{equation}
  \end{itemize}
  We show below that at this point all edge fluxes have been assigned to a cycle, and the remaining flux field is zero along all edges,
  \begin{equation}
    {\phi^i_j}^{(M+1)} = 0, ~\mathrm{for~all }~i,j.
   \label{eq:algo-terminate}
  \end{equation}

%%% -------------------------------------- %%
  \section{Existence of a valid decomposition}
  \label{sec:algo-proofs}
  The algorithm and its proof were first mentioned by MacQueen \cite{MacQueen1981} and later by Kalpazidou \cite{Kalpazidou1993}.
  We briefly review their argument.
  To show existence of the decomposition we demonstrate that for every flux field satisfying the steady-state condition, \eq{eq:node-condition}, the algorithm terminates with zero fluxes along all edges, (\ref{eq:algo-terminate}), and provides non-negative weights which fulfill the defining equation~(\ref{eq:weight-defn}).
  The algorithm always terminates in finite time because \(M\) is finite.
  Since the weight assigned to a cycle, \eq{eq:assigned-m}, is the minimum of all \({\phi^i_j}^{(k)}\) among the edges of cycle \(\alpha_k\), the new fluxes \({\phi^i_j}^{(k+1)}\) assigned by \eq{eq:assigned-phi} remain non-negative.
  Consequently, the steady-state weights \(m_{\alpha_{k}}^\ast\) are non-negative.

  We prove \eq{eq:algo-terminate} by contradiction.
  Suppose there is a flux
   \({\phi^i_j}^{(M+1)}\neq 0\).
  If this flux fulfills the node condition there is a cycle which could have been assigned a larger weight \(m_{\alpha_{k}}^\ast\), contradicting \eq{eq:assigned-m}.
  Hence, the remaining fluxes obey 
  \begin{equation}
   \sum_j \left({\phi^i_j}^{(M+1)} - {\phi^j_i}^{(M+1)}\right) \neq 0.
   \label{eq:algo-terminate-ic}
  \end{equation}
  In contrast, for every steady state the initial flux field \eq{eq:algo-initialize} fulfills the node condition (\ref{eq:node-condition}).
  Iterating the initial flux field we find
\begin{eqnarray}
    \nonumber
    0
    &=&
    \sum_j  \left({\phi^i_j}^{(k)} - {\phi^j_i}^{(k)}\right)
    \\ \nonumber
    &=&
    \sum_j  \left({\phi^i_j}^{(k+1)} - {\phi^j_i}^{(k+1)}\right) + m_{\alpha_{k}}^\ast \sum_j\left( {\chi^i_{j,\alpha_{k}}} - {\chi^j_{i,\alpha_{k}}} \right)
    \\    
    &=&
    \sum_j  \left({\phi^i_j}^{(k+1)} - {\phi^j_i}^{(k+1)}\right)
   \label{eq:algo-terminate-ind}
  \end{eqnarray}
  where we used \eq{eq:geometry-identity} in the last line.
  In contradiction to \eq{eq:algo-terminate-ic} this holds for every \(k\leq M\), proving \eq{eq:algo-terminate}.

  By construction the cycle fluxes obtained in this way fulfill \eq{eq:weight-defn}. We use \eq{eq:assigned-phi} and a telescope sum argument to obtain
  \begin{equation*}
   \sum_{k=1}^{M} m_{\alpha_k}^{\ast} {\chi^i_{j,\alpha_{k}}} \!
%   = \sum^M_{k=1}\left({\phi^i_j}^{(k)}-{\phi^i_j}^{(k+1)}\right) \!
   = {\phi^i_j}^{(1)} - {\phi^i_j}^{(M+1)} \!
   = {\phi^i_j}^*
  \end{equation*}
  where in the last equation we used the algorithm initialization \eq{eq:algo-initialize} and \eq{eq:algo-terminate}.

  \section{Number of cycles} \label{sec:numbers}
  Kalpazidou pointed out \cite{Kalpazidou1995} that the maximal number of cycles needed is the Betti number $M_B = \vert E \vert - N +1$ used in algebraic topology.
  This can also be seen directly from the algorithm.
 After all, in a worst-case scenario each non-trivial cycle reduces the flux along one edge.
  Because the iterated flux-field always fulfills the node condition (\ref{eq:node-condition}) the number of remaining edges at any vertex can never be exactly one.
  Then, as the graph is connected, at some point in the algorithm, $\vert E \vert - N$ nontrivial weights $m^*_\alpha$ have been assigned and the remaining flux field consists of $N$ nodes forming a cycle which will be assigned the last non-trivial weight.

  A minimal number of cycles with non-vanishing weights cannot be stated in general.
  However, symmetries present in the system that lead to the same fluxes at many edges may decrease this number as in the example below.

  The Betti number $M_B$ can also be related to the number of cycles used in SNT \cite{Schnakenberg1976}.
  In the latter theory, $U$ undirected edges yield a set of $M_{SNT}=U-N+1$ fundamental cycles.
  If the system exhibits only unidirectional edges, $M_B = M_{SNT}$.
  In the other limiting case dynamical reversibility holds and $M_B - M_{SNT} = U$.
  The $U$ additional numbers can be thought of as the detailed balance, \ie diffusive, part of the $U$ bidirectional transitions.
  Further, one can (by using the freedom of choice in the enumeration) specify a set of disjoint cycles to be part of the decomposition.
  A possible choice is to include the set of \(2\)-cycles (of which there are $U$ in a dynamically reversible system).
  The result is a splitting of the fluxes in a detailed-balance part (the set represented by the \(2\)-cycles), and the remaining current part.
  This resembles the approach in \cite{Zia+Schmittmann2007}, but is conceptionally different because the decomposition here does not discard the information stored in the \(2\)-cycles.

%% -------------------------------------- %%
\section{Detailed balance dynamics on cycle space} \label{sec:detbal}
%% -------------------------------------- %%
  \begin{figure*}
    \hfill
    \includegraphics[width = .3 \textwidth]{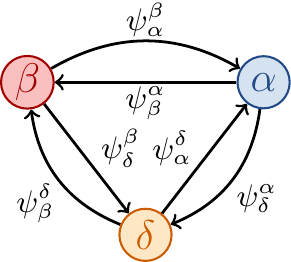}\hspace{0.05\textwidth} 
    \hfill
    \includegraphics[width= .3\textwidth]{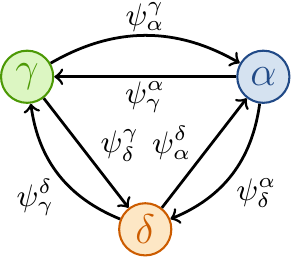} \vspace{-.015\textwidth}
    \hfill \ 
   \caption{ \label{fig:transformed} Transformed graph \(H\) obtained for the original graph \(G\) for the two decomposition of the flux field introduced in figure \ref{fig:markov-graph}.
   }
  \end{figure*}
%% -------------------------------------- %%

  The set of weights
   \(\{ m_{\alpha_k}^{\ast} \}\)
  can be interpreted as a mapping that transforms the original graph \(G=(V,E)\) into a new one \(H = (C,E_C)\), see figure \ref{fig:transformed}.
  For instance, the vertex \(\alpha \in C\) represents the cycle \(\alpha\) in \(G\) with the non-zero weight \(m_\alpha^{\ast}\) as identified by the algorithm.
  A directed edge
   \((\alpha,\beta)\in E_C\)
  indicates that two cycles share at least one vertex of \(G\), \ie one state of the original system.
  Each edge \((\alpha,\beta)\) of the transformed graph is associated with a transition rate \(b^\alpha_\beta\).
  In the analogy of the mass transit system
   \(\psi^\alpha_\beta := m^*_\alpha b^\alpha_\beta\)
  characterizes the number of passengers changing from line \(\alpha\) to line \(\beta\) in the stationary system.

  We shall call the operation \(G \rightarrow H\) the {\it cycle transform}.
By virtue of (\ref{eq:weight-defn}) the steady-state fluxes can be calculated from $\left\{ m_\alpha^* \right\}$ and $\{ \chi^\alpha_\beta \}$.
If the steady-state distribution $\{p^*_i\}$ is known, the full Markovian dynamics on the original state space can be reconstructed.
In terms of cycles, the $\{p^*_i\}$ can be interpreted as loops associated with each vertex in $G$ (\cf the discussion of the discrete case below).

  To find the rate constants \(b^\alpha_\beta\) we realize that in the steady state at each vertex \(v_i\) (\ie station, in the socio-physical picture) a constant number of passengers arrives per unit time.
  This number is proportional to the overall influx
   \(\sum_\gamma \chi_{i,\gamma}m^*_\gamma=\sum_{j\neq i} {\phi^j_i}^*\).
  The passengers carry tickets indicating which line they are running on.
  Upon arrival at the station, a passenger enters his ticket into a ticket machine that provides him with a new one.
  The probability to draw a ticket for line \(\beta\) is given by ratio of the weight of line $\beta$ to the weights of all lines serving station $i$, \ie
  \begin{equation}
   b_\beta^{(i)}=\frac{m^*_\beta}{\sum_\gamma \chi_{i,\gamma}m^*_\gamma}.
   \label{eq:bi-defn}
  \end{equation}
  The total flux \(\psi^\alpha_\beta\) from line \(\alpha\) to line \(\beta\) is obtained by summing the local exchange flux
   \(m^*_\alpha b_\beta^{(i)} \)
  over all mutual stations where
   \(\chi_{i,\beta}\chi_{i,\alpha}=1\)
  \begin{equation}
   \psi^\alpha_\beta = \sum_{i}\chi_{i,\beta}\chi_{i,\alpha} m_\alpha^{\ast} b_\beta^{(i)} = m^*_\alpha \sum_i \frac{\chi_{i,\beta}\chi_{i,\alpha}}{\sum_\gamma \chi_{i,\gamma} m^*_\gamma }m^*_\beta = \psi^\beta_\alpha.
   \label{eq:transformed-flux}
  \end{equation}
  A remarkable feature of this new formulation is that the cycle-space fluxes fulfill detailed balance (\(\psi^\alpha_\beta = \psi^\beta_\alpha~\mathrm{for~all }~\alpha,\beta\)).
  In the steady state this is a microscopically balanced ticket exchange.
  It means, that on average, passengers arriving at a station just exchange tickets with other passengers, and board the line for which their new ticket holds.

  Because of detailed balance in \(H\) we can proceed along the line indicated by (\ref{eq:Potential-DB}).
  Replacing \(w^i_j\) by \(b^\alpha_\beta\), one obtains a potential \(\mathcal{H}_\alpha\), such that the occupation numbers \(m_\alpha^{\ast}\) are given by Boltzmann weights,
  \begin{equation}
   m_\alpha^{\ast} = \mathcal Z^{-1} \exp(-\mathcal{H}_\alpha).
   \label{eq:BoltzmannWeights}
  \end{equation}
  Here the partition function 
  \begin{equation}
   \mathcal Z = \sum_{\alpha} \tau_\alpha \exp\left( -\mathcal H_\alpha \right) \, ,
   \label{eq:partition-function}
  \end{equation}
  includes the average cycle period 
  $\tau_\alpha = \sum_i \chi_{i,\alpha}\langle \tau_i \rangle$.
%  \begin{equation}
%   \tau_\alpha = \sum_i \chi_{i,\alpha}\langle \tau_i \rangle.
%   \label{eq:cycle-periods}
%  \end{equation}
  After all, the weights \(m^*_\alpha\) are not probabilities.
  According to equations~(\ref{eq:transition-matrix}), (\ref{eq:flux-defn}) and (\ref{eq:weight-defn}) they rather fulfill
  \begin{equation*}
    \sum_{\alpha} m^*_{\alpha} \tau_\alpha = \sum_{\alpha} m^*_{\alpha} \sum_i \chi_{i,\alpha} \langle \tau_i \rangle = \sum_i  p^*_i = 1.
  \end{equation*}
  In summary, the potential \(\mathcal{H}_\alpha\) is obtained from the NESS fluxes ${\phi^i_j}^*$ by determining the population density \(m_\alpha^{\ast}\) of the cycles, followed by equations~(\ref{eq:bi-defn}), (\ref{eq:transformed-flux}) and finally (\ref{eq:Potential-DB}). Though our approach does not require knowledge of the invariant measure $p_i^*$, the steady-state fluxes ${\phi^i_j}^*$ have to be known. 
  From an analytical point of view this requires the full solution of the mathematical problem.
  However, in experiments fluxes might be easier accessible than probabilities.

  \section{Averages on cycle space} \label{sec:averages}

  For every well-defined mapping
   \(F: \alpha \mapsto F_\alpha\)
  from the set of cycles to the real numbers we define the cycle average
  \begin{equation}
   \langle F \rangle_{C} := \sum_{\alpha} m_{\alpha}^{\ast} F_\alpha.
   \label{eq:cycle-average}
  \end{equation}
  For instance, for the characteristic functions \(\chi^i_{j,\alpha}\) we have
   \( \langle \chi^i_j \rangle_C = {\phi^i_j}^* \)
  by eqs.~(\ref{eq:weight-defn}, \ref{eq:cycle-average}).
  On the other hand,
   \(\langle 1 \rangle_C \neq 1\),
  because the edge fluxes are not normalized weights.
  
  Now let us consider cycle-space observables related to physical quantities.
  Consider some matrix
   \(F\in \mathbb {R}^{N\times N}\).
  We can interpret this quantity as the change of some physical observable due to the transitions between different states. 
  We define 
  \begin{equation}
   J_{F}(t)=\sum_{i,j} F^i_j \phi^i_j(t) =: \langle F \rangle_{2,t}
   \label{eq:macro-fluxes}
  \end{equation}
  as the average flux of quantity \(F\) at time \(t\).
  The last equivalence is the definition of the average as the two-point probability-density function (propagator) at time \(t\).
  For antisymmetric \(F\) one has
   \(J_F = 1/2 \sum_{i,j} F^i_j I^i_j\).

  To connect this with the cycle transform we define an observable
  \begin{equation}
   F_\alpha = \sum_{i,j}\chi^i_{j,\alpha}F^i_j
   \label{eq:F-cyclespace}
  \end{equation}
  which is the integrated contribution of \(F\) along cycle \(\alpha\).
  With the linearity of the averages
  \begin{equation}
    J_F^*=\lim_{t\to\infty}\langle F \rangle_{2,t} = \sum_{i,j} F^i_j {\phi^i_j}^* = \sum_\alpha \sum_{i,j} m_\alpha^* \chi^i_{j,\alpha} F^i_j = 
    % \sum_\alpha m_\alpha^* F_\alpha = 
    \langle F \rangle_{C}.
   \label{eq:average-equivalence}
  \end{equation}

%------------------------------------------
\begin{table}
  \centering
  \begin{tabular}{|c|c||c|c|}
    \hline \hline
    \bf state & \bf  configuration & \bf state & \bf configuration\\
    \hline \hline
    1 & \X\O\X\O & 2 & \O\X\X\O \\
    3 & \X\O\O\X & 4 & \X\X\O\O \\
    5 & \O\O\X\X & 6 & \O\X\O\X \\
    \hline
  \end{tabular}
  \caption{The six possible configurations for the TASEP example. The corresponding network of states is shown in figure~\ref{fig:TASEP-example}. Transitions $5\rightarrow 1$ and $6\rightarrow 4$ involve a particle leaving at the right site and entering at the left site.} 
  \label{tab:TASEP-states}
\end{table}
%------------------------------------------

%% -------------------------------------- %%
\section{Dynamical reversibility and non-equilibrium thermodynamics}\label{sec:physics}

  Here we provide the connection of averages in the general formalism to the ones needed to describe physical currents in non-equilibrium systems.
  To that end we consider dynamically reversible systems, (\ref{eq:dynamical_reversibility}).
  This is no constraint because in physical systems one has reversible microscopic laws.
  This means that for every microscopic ``forward'' trajectory leading the system from state \(i\) to \(j\) also the time-reversed ``backward'' trajectory from \(j\) to \(i\) is a solution of the equations of motion.
  Remember that this is not needed for the application of the cycle transform, as the example of figure \ref{fig:markov-graph} shows. 

  Dynamical reversibility allows the connection of Markov process to (non-equilibrium) thermodynamics \cite{Penrose1970,Schnakenberg1976,Hill1977,Esposito+vdBroeck2010}.
  The central quantities describing a NESS are the non-zero macroscopic currents \(I\) which are driven by macroscopic affinities \(A\).
 One can consistently define them also on the level of stochastic transitions:
\begin{subequations}
    \begin{eqnarray}
     I^i_j &%
     :=& \phi^i_j - \phi^j_i, \\%
     [2mm] A^i_j &%
     :=&
     \log\phi^i_j-\log\phi^j_i.
    \end{eqnarray}%
    \label{eq:thermo_defn}%
\end{subequations}
  Further, a connection with entropy production and therefore heat dissipation can be made (\cf also the analogies given in \ref{sec:analogy}).
   Observe that
    \(\mathrm{sgn}(I^i_j) = \mathrm{sgn}(A^i_j)\).
   Consequently, the positive total entropy production can always be expressed \cite{Schnakenberg1976} as
   \begin{equation}
    P_{tot} = \frac 1 2 \sum_{i,j} A^i_j I^i_j.
    \label{eq:entropy-production}
   \end{equation}
   {\it Cycle affinities} are the integrated values \eq{eq:F-cyclespace} of the anti-symmetric affinity matrix $A^i_j$.
   They are related to the macroscopic thermodynamic affinities as was first realized by Hill \cite{Hill1977} and formulated somewhat differently by Schnakenberg \cite{Schnakenberg1976}.
   With the decompositions introduced above one generalizes the results of Schnakenberg \cite{Schnakenberg1976} and the Quians \cite{Jiang_etal2004}
  \begin{equation}
   P^*_{tot} = J^*_{A} = \langle A \rangle_{C} = \sum_{\alpha} m_{\alpha}^{\ast} A_\alpha
   \label{eq:entropyProductionAverage}
  \end{equation}
  for the entropy production in the steady state to cycles obtained by the deterministic algorithm presented above.
  In the context of entropy production cycles are also used \cite{Andrieux+Gaspard2007,Faggionato+dPietro2011,Jiang_etal2004} for the well-known fluctuation relations for the entropy production along (a set of) individual random trajectories (\cf \cite{Lebowitz+Spohn1999,Seifert2005}).

\section{Change of dominant paths: 2-particle 4-site driven TASEP} \label{sec:example}
\begin{figure*}
%\centering
%\subfloat[Transition rates of the TASEP example with a variable jump rate $x$ for transitions involving a particle jump over the boundary.]{
%\label{fig:TASEP-example} %% label for first subfigure
\hfill
\includegraphics[width=0.36\textwidth]{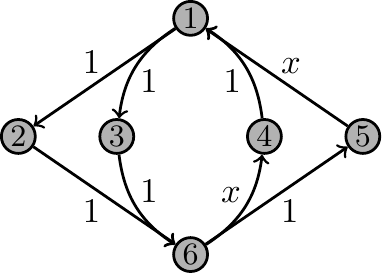}
\hfill
%}%
% \hspace{.05 \textwidth}
%%—-start of second subfigure—-
%\subfloat[Dependence of the steady-state fluxes on $x$. For clarity, the edge fluxes shown are divided by the common factor $x/C(x)$, \cf (\ref{eq:TASEP_example_fluxes}).]{
%\label{fig:TASEP-example:fluxes} %% label for second subfigure
\includegraphics[width=0.36\textwidth]{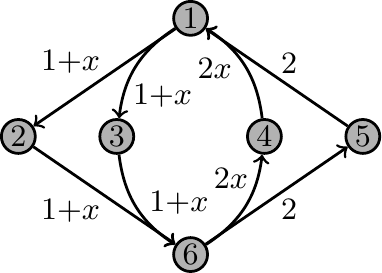}
\hfill
\ 
%}%
\caption{The network of states for the TASEP example. The rates shown in (a) lead to the steady state shown in (b). For $x=2$ the fluxes are proportional to the ones shown in the original example, figure \ref{fig:markov-graph}. 
(a) Transition rates of the TASEP example with a variable jump rate $x$ for transitions involving a particle jump over the boundary.
(b) Dependence of the steady-state fluxes on $x$. For clarity, the edge fluxes shown are divided by the common factor $x/C(x)$, \cf (\ref{eq:TASEP_example_fluxes}).
}
\label{fig:TASEP-example} %% label for entire figure
\end{figure*}

In this section we illustrate the consequences of a parameter change on the selection of paths in a variant of a totally asymmetric exclusion process (TASEP), \cite{Derrida1998}.
Consider a one-dimensional periodic lattice (\ie a ring) with four sites.
On the lattice we put two particles and allow them to move in only one direction.%
\footnote{Note that though the process is physically motivated, the system is lacking dynamic reversibility.}
Each site can only be occupied by one particle so the particles are not independent of each other.
The rates for particles jumping from one site to the next are all equal (and set to unity) but one, which is set to a value $x>0$.
Particles are accelerated or slowed down at that site.
The system's state is represented by binary $4$-tupels with ``\X'' representing an occupied and ``\O'' an empty site, see table \ref{tab:TASEP-states}.
Figure \ref{fig:TASEP-example}(a) shows the network of states with its positive transition rates.
The rate for a particle jumping over the edge from the last to the first site has magnitude $x$.
This corresponds to transitions $5\rightarrow 1$ and $6\rightarrow 4$, as they are the ones utilizing the periodic boundary conditions, \cf table \ref{tab:TASEP-states}.

The steady-state distribution is
\begin{eqnarray*}
  p &=& \left(x(1+x),x(1+x),x(1+x),2x^2,2,2x\right)/C(x),
  \\ 
  C(x) &=& 2 + 5x + 5x^2  
\end{eqnarray*}
leading to the steady-state fluxes
\begin{eqnarray}
    \phi^1_2 = \phi^2_6 = \phi^1_3 = \phi^3_6 &=&\left(x/C(x)\right) (1+x),\nonumber \\ 
    \phi^6_4 = \phi^4_1  &=&\left(x/C(x)\right)2x,\nonumber \\
    \phi^6_5 = \phi^5_1 &=&\left(x/C(x)\right)2.  
    \label{eq:TASEP_example_fluxes}
\end{eqnarray}
For $x=2$ the fluxes are the ones of the initial example (figure \ref{fig:markov-graph}) up to a factor of $2/C(2)=\frac{1}{16}$.

%------------------------------------------
%------------------------------------------
\begin{table}
  \centering
  \begin{tabular}{|c|c|c|c|}
    \hline \hline
    \bf cycle & \bf sequence & \bf gait & \bf graph\\
    \hline \hline
    $\alpha$ &  \X\O\X\O $\rightarrow$\X\O\O\X $\rightarrow$\O\X\O\X $\rightarrow$\X\X\O\O \quad(\small$1\rightarrow3\rightarrow6\rightarrow4$) & 1f &\includegraphics[height = 6mm]{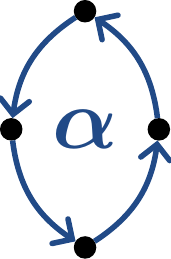}\\
    $\beta$ & \X\O\X\O $\rightarrow$\O\X\X\O $\rightarrow$\O\X\O\X $\rightarrow$\O\O\X\X \quad(\small$1\rightarrow2\rightarrow6\rightarrow5$)& 1b &\includegraphics[height = 6mm]{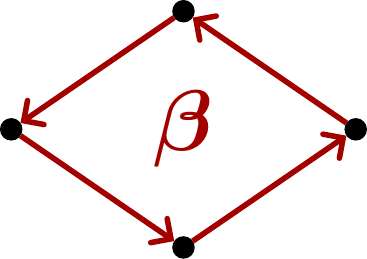}\\
    $\gamma$ &\X\O\X\O $\rightarrow$\X\O\O\X $\rightarrow$\O\X\O\X $\rightarrow$\O\O\X\X \quad(\small$1\rightarrow3\rightarrow6\rightarrow5$)& 2f &\includegraphics[height = 6mm]{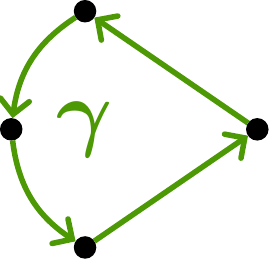}\\
    $\delta$ & \X\O\X\O $\rightarrow$\O\X\X\O $\rightarrow$\O\X\O\X $\rightarrow$\X\X\O\O \quad(\small$1\rightarrow2\rightarrow6\rightarrow4$)& 2b &\includegraphics[height = 6mm]{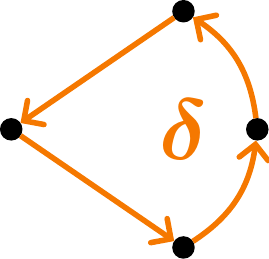}\\
    \hline
  \end{tabular}
  \caption{The four different cycles of the TASEP example. Each cycle corresponds to a different ``gait'' of the two particles. Gaits are characterized by step size (1 or 2) and  whether the front or back particle moves first (f or b).}
  \label{tab:cycle-gaits}
\end{table}
%
%------------------------------------------
\begin{table}
  \centering
  \begin{tabular}{|c|c|c|}
    \hline \hline
    \bf region & \bf fluxes & \bf decompositions\\
    \hline \hline
    \multirow{2}{*}{$x>1$} &\multirow{2}{*}{ $2<1+x<2x$} & $(x+1){\boldsymbol \alpha} + (x-1){\boldsymbol \delta} +2{\boldsymbol \beta}$,\\
    & & $(x+1){\boldsymbol \delta} + (x-1){\boldsymbol \alpha} +2{\boldsymbol \gamma}$\\
      \hline
      $x=1$ & $2=1+x=2x$ & $2{\boldsymbol\alpha} + 2{\boldsymbol \beta} $,\quad $2{\boldsymbol \gamma} + 2{\boldsymbol \delta}$\\
    \hline
    \multirow{2}{*}{$x<1$} &\multirow{2}{*}{ $2>1+x>2x$} & $(x+1){\boldsymbol \beta} + (1-x){\boldsymbol \gamma} +2x{\boldsymbol \alpha}$,\\
    & & $(x+1){\boldsymbol \gamma} + (1-x){\boldsymbol \beta} +2x{\boldsymbol \delta}$\\
    \hline
  \end{tabular}
  \caption{Different decompositions depending on $x$. A transition happens at $x=1$. The numerical values shown are the weights divided by the common factor $x/C(x)$.}
  \label{tab:TASEP-decompositions}
\end{table}
%------------------------------------------

We now take a closer look at the four cycles present in the system.
The cycles correspond to four different gaits of the particles characterized by the step size and whether the front or back particle particle moves first.
The distinction between front and back is arbitrary, due to the periodicity of the system.
Particles cannot overtake, so one can distinguish particles.
In this case, completing any of the cycles leads to a state that has the particles switched.
Completing two cycles would then bring the system to its original configuration.
The full characterization is shown in table \ref{tab:cycle-gaits}.

One can easily check that the algorithm given in section \ref{sec:transform} leads to only two possible decompositions for any positive $x$.
As said above, which of those two decompositions is realized depends on the ordering of cycles.
Further one can see that there are three regions for $x$ corresponding to qualitatively different decompositions (see table~\ref{tab:TASEP-decompositions}).
For $x>1$ we always end up with non-zero weights for one of the decompositions shown in figure~\ref{fig:markov-graph}, which are $\{\alpha,\gamma,\delta\}$ or $\{\alpha,\beta,\delta\}$.
For $0<x<1$ we have a qualitatively different behavior as the decomposition will either feature $\{\alpha,\beta,\gamma\}$ or $\{\beta,\gamma,\delta\}$.
At the transition point $x=1$ the possible decompositions,$\{\alpha,\beta\}$ or $\{\gamma,\delta\}$, consist of two cycles.

We now look at the number of cycles needed for the decomposition and compare it to the cycles used in other theories.
The Betti number for this system is $M_B = 8 - 6 + 1 = 3$.
It agrees with the number of fundamental cycles used in SNT,  $M_{SNT} = M_B$, because we have only uni-directional transitions.
In the highly symmetric case $x=1$ our algorithm yields a smaller number of cycles, $2<3=M_B$.
Table \ref{tab:TASEP-decompositions} summarizes the decompositions and numerical values of the weights in the three regions.

The crucial point is that the decomposition structure changes at the transition point $x=1$, meaning that some zero-weights suddenly become positive while other go to zero.
We can also note this by only looking at one cycle, which we fix to be the first one to be considered by the algorithm, \ie $\alpha_1$ .
By that, its appearance in the decomposition is ensured.
Then, at the transition point $x=1$ we observe a discontinuous change in the derivative $\rmd m_{\alpha_1}/\rmd x$ of its weight from $1$ to $2$.

These discontinuous changes in the structure of cycles and their weights are related to a change in the dominant paths in the network of states.
In our example, this is triggered by the change from accelerating to decelerating a particle when it crosses the periodic boundary.

The ASEP with a modified transition (or bond) rate on a periodic one-dimensional lattice has been introduced in \cite{Janowsky+Lebowitz1994}, where it was observed that a fast  bond leads only to local correlations, whereas a slow bond can have long range effects (due to particles piling up).
An exact formula for the stationary measure of the slow bond system remains an open problem, and it would be interesting to investigate possible connections with the changes in cycle structures.

%% -------------------------------------- %%
  \section{Conclusion and outlook}\label{sec:conclusion}

  In this work we presented a mapping, the cycle transform, that generally applies to steady states of Markov processes on a finite state space.
  It can be used to transform a non-equilibrium steady state represented by a graph \(G\) into an equilibrium steady state on a graph \(H\) whose vertices are appropriately chosen cycles in \(G\).

The presented mapping is obtained by using a deterministic algorithm rather than a stochastic algorithm \cite{Kalpazidou2006,Jiang_etal2004}.
Therefore the theory lies between theories based on all possible flux cycles (\cf Hill's theory, \cite{Hill1977}) and theories using fundamental current cycles (\cf SNT, \cite{Schnakenberg1976,Andrieux+Gaspard2007}).
  The non-uniqueness of our decomposition can be used to separate detailed balance contributions (\(2\)-cycles) from non-equilibrium currents (non-trivial cycles).

  Further, the connection between averages defined on the space of cycles to steady-state averages was made.
  For physical systems, a natural symmetry on \(G\), called dynamical reversibility, allows us to relate the method to currents in non-equilibrium thermodynamics.
  
  The suggested approach also has interesting parallels to the theory of dynamical systems, especially chaos theory \cite{Cvitanovic+Rondoni2010,Jiang_etal2004}.
  In chaos theory cycles, \ie unstable periodic orbits of the dynamical system, play a crucial role.
  They lie dense in phase space such that trajectories can be seen as a realization of a random-walk dynamics between cycles, 
  similar to the dynamics in cycle space considered in the present study.
  Expectation values in such systems can also be calculated using cycle expansions.

  Finally, we illustrated the method by exploring a TASEP example where one can interpret the cycles as different periodic gaits.
  It exhibits a crossover of the preferred paths in response to a parameter change.
  This is reflected in a discontinuous change of weights of the cycles.
  In addition, there is a topological change:
  At the transition point, the structure of the cycle decomposition changes.

  In forthcoming work, the cycle transform might serve as another perspective on thermodynamic machines where different cycles represent the different operation modes.
  A well-studied example is the steady-state dynamics of the molecular motor kinesin \cite{Liepelt+Lipowsky2009}.
  For such small machines thermal fluctuations play a crucial role.
  The cycle-transform representation of the entropy production, eq.~(\ref{eq:entropyProductionAverage}), is an important perspective to this problem.
  Cycle affinities and cycle currents can be used to formulate fluctuation relations \cite{Lebowitz+Spohn1999,Andrieux+Gaspard2007,Faggionato+dPietro2011}.

%% -------------------------------------- %%
\begin{acknowledgments}

  The authors are indebted to L. Rondoni, M.~Brinkmann, R. Lipowsky,
  P. Cvitanovi\'c, U. Seifert, B.~Drossel, M. Denker, A.~Fingerle, and
  V. Zaburdaev for inspiring discussions and helpful hints.

\end{acknowledgments}

\appendix

  \section{An electric and thermodynamic analogy}\label{sec:analogy}
  \begin{table}
   \centering
   \begin{tabular}{|c|c|cc|}
    \hline \hline {\bf symbol}&%
    {\bf analogy}&%
    {\bf thermodynamic}&%
    {\bf electric}\\
    \hline \hline \(V_i\) &%
     \( -\log p_i\) &%
    \multicolumn{2}{c|}{\small potential} \\
    \hline \(U^i_j\) &%
     \( \log [p_i/p_j]\)&%
    {\small total differential}&%
    {\small voltage}\\
    \hline \(I^i_j\) &%
     \(\phi^i_j - \phi^j_i\) &%
    \multicolumn{2}{c|}{\small current} \\
    \hline \(A^i_j\) &%
     \(\log[\phi^i_j / \phi^j_i]\) &%
     {\small affinity, \it {gross FED} }&%
    - \\
    \hline \(\mathcal{E}^i_j\) &%
     \(\log[w^i_j / w^j_i]\) &%
     {\it basic FED}&%
    {\small electromotance}\\
    \hline \(R^i_j\) &%
     \( U^j_i/I^i_j\) &%
    -&%
    {\small resistance}\\
    \hline \(P_{sys}\) &
     \(\frac 1 2 \sum_{i,j}U^i_j I^i_j\)
    &%
    {\small system entropy change}&%
    {\small power}\\
    \hline
   \end{tabular}
   \caption{Electric and thermodynamic analogies. FED denotes free energy differences as in Hill's theory, \cite{Hill1977}.}
   \label{tab:analogies}
  \end{table}

  In this section we introduce an analogy relating Markov processes, thermodynamics and electrical circuits.
  Different electric analogies have been presented in the literature that are suitable for different purposes (see e.g.
  \cite{Zia+Schmittmann2007,Feller1968}).
  Hill also noticed the connection of the logarithmic ratios of fluxes and transition rates with differences of free energies \cite{Hill1977}.
  The appropriate analogies are summarized in table \ref{tab:analogies}.
  In this analogy the quantities defined above have the properties of their electrical counterparts: \\
   \(U\), \(I\), \(A\) and \(\mathcal E\) are asymmetric and the resistance \(R\) is symmetric and positive.
  The definition of the fluxes, \eq{eq:flux-defn}, obeys Kirchhoff's equation \cite{Kirchhoff1847},
  \begin{equation}
   U^i_j + \mathcal{E}^i_j = R^i_j I^i_j , 
   \label{eq:Kirchhoff}
  \end{equation}
  which states 
  that if no current is flowing between two nodes with a battery-like element connecting them, a voltage difference \(U\) is created.
  This voltage is the negative of the electromotance \(\mathcal E\) of the battery.
  However, if a current is running over a resistor \(R\), it obeys an Ohmic law and the voltage drops by \(R\cdot I\).
  Kirchhoff's current law (``node rule'') amounts to \eq{eq:node-condition}.
  Kirchhoff's voltage law (``mesh rule'') states that integrating the voltage differences around a closed cycle is zero.
  This also holds in our analogy.
  It is the basis for the identification of \(U\) with a total differential in thermodynamics.
  
  Finally, the quantity \(P_{sys}\) describes the change of the system's Gibbs entropy
   \(S_{sys} := -\sum_i p_i \log p_i\)
  as the systems undergoes its dynamics, 
  \begin{equation}
   P_{sys} = \frac{\rmd}{\rmd t}S_{sys}.
   \label{eq:entropy-variation}
  \end{equation}
  It vanishes in the steady state, and can be related to the irreversible entropy production \(P_{tot}\) by defining an entropy flux to the medium \cite{Schnakenberg1976,Zia+Schmittmann2007}
  \begin{equation}
   P_{med} = \frac 1 2 \sum_{i,j} \left( \phi^i_j - \phi^j_i \right) \log \frac{w^i_j}{w^j_i} \, .
   \label{eq:entropy-flux}
  \end{equation}
  One then finds
   \(P_{tot}=P_{sys}+P_{med}\).
  Introducing thermodynamic analogues one obtains
   \(P_{med} = \frac 1 2 \sum_{i,j} I^i_j \mathcal E^i_j\)
  such that 
  \mbox{\(P_{tot} =\frac 1 2 \sum_{i,j} I^i_j ( U^i_j+\mathcal E^i_j)\).}
  Hence, the definitions of table \ref{tab:analogies} are consistent with the definitions made earlier, and
   \(A^i_j = U^i_j+\mathcal E^i_j\).

  The analogy is not perfect, however.
  Consider a simple cycle with the same current flowing through all nodes.
  Then the potential difference between two non-adjacent nodes \(i\) and \(j\) cannot be obtained from an effective resistance (or electromotance) which is the sum of the individual resistances (or electromotances) of the edges connecting \(i\) to \(j\) as it would be the case in an electrical network.
%% -------------------------------------- %%

\section{Discrete case} \label{sec:discrete}

 If time is measured in discrete units $\tau$ one obtains a Markov chain.
 In that case one has transition probabilities \(0\leq a^i_j \leq 1\) rather than transition rates \(w^i_j\). 
 Instead of a waiting time \(\tau_i\) one has a staying probability \(a^i_i\neq 0\). 
 The transition matrix is \((A)^i_j \equiv a^i_j\) and the evolution of the probability distribution \(p_i\) obeys
 \begin{equation}
   p_i(t+1) = \sum_j a^j_i p_j(t).
   \label{eq:discrete-evolution}
 \end{equation}
 With the normalization for the transition probabilities 
 \begin{equation}
   \sum_j a^i_j = 1
   \label{eq:jump-normalization}
 \end{equation}
 and defining discrete time fluxes (\ie joint probabilities) \(\phi^i_j(t)=p_i(t) a^i_j\) one can rewrite
 \eq{eq:discrete-evolution} into a master equation
 \begin{equation}
   p_i(t+1)-p_i(t) = \sum_{j\neq i} \phi^j_i -\phi^i_j.
   \label{eq:master-discrete}
 \end{equation}
 The steady-state condition is formally identical to \eq{eq:node-condition}.
 Therefore, all relations for the cycle representation also hold in the discrete case.
 \\
 Further, the analogies presented in table \ref{tab:analogies} hold if one substitutes the transition rates \(w^i_j\) for the jump probabilities \(a^i_j\).
 Still, there is a subtle difference we would like to point out:\\
 The cycle transform introduced above only uses fluxes \(\phi^i_j\) with \(i\neq j\). 
 Therefore, the number of variables to be specified is \(N(N-1)\).
 In the discrete case, it is straightforward to include the disjoint {\it loop fluxes} \(\phi^i_i\) into the cycle transform by specifying \(N\) additional variables.
 One can then uniquely reconstruct the transition matrix \(A\) and the steady state probabilities \(p_i^*\) from the fluxes by using
 \eq{eq:discrete-evolution} and the definition of the fluxes.
 As in the continuous-time case, to reconstruct the full steady state one has to specify \(N\) additional variables that do not directly influence the cycle transform algorithm.

%%%%%%%%%%%%%%%%%%%%%%%%%%%%%%%%%%%%%%%%%%%%%%%%%%%%%%%%%%%%%%
% \bibliographystyle{apsrev4-1}
% \bibliography{cycles}
%%%%%%%%%%%%%%%%%%%%%%%%%%%%%%%%%%%%%%%%%%%%%%%%%%%%%%%%%%%%%%
%merlin.mbs apsrev4-1.bst 2010-07-25 4.21a (PWD, AO, DPC) hacked
%Control: key (0)
%Control: author (72) initials jnrlst
%Control: editor formatted (1) identically to author
%Control: production of article title (-1) disabled
%Control: page (0) single
%Control: year (1) truncated
%Control: production of eprint (0) enabled
%

 \end{document}